\documentclass[twocolumn,groupedaddress,showpacs,nofootinbib]{revtex4-1}

\usepackage[colorlinks,pdfstartview=FitH]{hyperref}
\hypersetup{linkcolor=blue,citecolor=blue,filecolor=black,urlcolor=blue}

\usepackage{amsmath,amssymb,bm}
\usepackage{graphicx}
\usepackage{amsfonts}
\usepackage{color}

\usepackage{amsfonts}
\usepackage{bbm}
\usepackage{latexsym}
\usepackage{dcolumn}

\usepackage{natbib}
\usepackage{twoopt}
\begin{document}

\baselineskip=15pt \parskip=5pt

\vspace*{3em}

\title{Possible s-wave annihilation for MeV dark matter with the 21-cm absorption}

\author{Lian-Bao Jia }
\email{jialb@mail.nankai.edu.cn}
\affiliation{School of Science, Southwest University of Science and Technology, Mianyang
621010, China}

\author{Xu Liao }
\affiliation{School of Science, Southwest University of Science and Technology, Mianyang
621010, China}

\begin{abstract}

The CMB observation sets stringent constraints on MeV dark matter (DM) annihilating into charged states/photons in s-wave, and the recent observation of the 21-cm absorption at the cosmic dawn reported by EDGES is also very strict for s-wave annihilations of MeV DM. The millicharged DM with p-wave dominant annihilations during the freeze-out period are considered in literatures to give an explanation about the 21-cm absorption, with photon mediated scattering cooling the hydrogen. In this paper, we focus on the annihilation of millicharged DM being s-wave dominant. To explain the 21-cm absorption and meanwhile be compatible with the CMB and 21-cm absorption bounds on DM annihilations, we consider the annihilation close to the resonance, with the new mediator (here is dark photon) mass being slightly above twice of the millicharged DM mass. In this case, the annihilation cross section at the temperature $T \to 0$ could be much smaller than that at $T_f$, which would be tolerated by the bounds on DM annihilations, avoiding the excess heating from DM s-wave annihilations to the hydrogen gas. The beam dump and lepton collider experiments can be employed to hunt for millicharged DM via the production of the invisible dark photon.

\end{abstract}

\maketitle

\section{Introduction}

For dark matter (DM) particles with masses in a range of ten MeV to hundreds TeV, the relic abundance of DM can be obtained via the thermal freeze-out of DM. One DM candidate extensively concerned is weakly interacting massive particles (WIMPs) with masses in GeV-TeV scale, and results from recent DM direct detections \cite{Agnese:2017jvy,Akerib:2016vxi,Cui:2017nnn,Aprile:2018dbl,Akerib:2017kat,Xia:2018qgs,Aprile:2019dbj,Amole:2019fdf} set stringent constraints on WIMP-nucleon scatterings. In case of DM being lighter and in MeV scale, the MeV DM could evade the DM-target nucleus scattering hunters. Thus, the MeV DM is of our concern.

The bulk of the cosmological matter density (about 84\%) is contributed by DM \cite{Aghanim:2018eyx}, and the typical annihilation cross section of DM during the freeze-out period is about $3 \times 10^{-26} \mathrm{cm}^3/\mathrm{s}$. Furthermore, the cosmic microwave background (CMB) observations at the recombination epoch set upper limits on s-wave annihilations of MeV DM with the annihilation products being of charged states/photons \cite{Aghanim:2018eyx,Slatyer:2015jla}, which are much below the annihilation cross section required by the relic abundance of DM. In addition, the constraint from the recent observation of the 21-cm absorption \cite{Bowman:2018yin} at the cosmic dawn is also very strict for the s-wave annihilation of MeV DM \cite{DAmico:2018sxd,Cheung:2018vww,Liu:2018uzy}, as the energy injection from DM s-wave annihilations would heat the hydrogen gas. Therefore, the MeV DM with p-wave dominant annihilations during the freeze-out period are generally considered in literatures \cite{McDonald:2000bk,Diamanti:2013bia,Jia:2016uxs}.

Here we will focus on the 21-cm absorption. The enhanced 21-cm absorption observed by the EDGES Collaboration \cite{Bowman:2018yin} indicates that the neutral hydrogen at the cosmic dawn would be colder than expected, and a feasible mechanism is that the hydrogen is cooled by the scattering with MeV millicharged DM,\footnote{A millicharge may be from a kinetic mixing of an extra massless gauge boson \cite{Holdom:1985ag}, or other scenarios, see e.g., Refs. \cite{Kors:2004dx,Feldman:2007wj,Cheung:2007ut,Cline:2012is,Kouvaris:2013gya}.} with photon being the mediator in the scattering \cite{Barkana:2018lgd,Xu:2018efh,Munoz:2018pzp,Fialkov:2018xre,Mahdawi:2018euy,Boddy:2018wzy}.\footnote{See Refs. \cite{Mirocha:2018cih,Li:2018kzs,Feng:2018rje,Fraser:2018acy,Pospelov:2018kdh,Widmark:2019cut} for more about the 21-cm absorption.} For the 21-cm brightness temperature $T_{21} = - $300 mK (the upper limit from EDGES), the required MeV millicharged DM is in a mass range about 10$-$35 MeV, which carries a millicharge $\eta e$ with $\eta$ $\sim$ $5 \times 10^{-6} - 5 \times 10^{-5}$, and the millicharged DM makes up a small fraction $f_\mathrm{DM}$ of the total DM relic density \cite{Liu:2018uzy,Berlin:2018sjs,Barkana:2018qrx,Slatyer:2018aqg,Munoz:2018jwq,Kovetz:2018zan}, i.e., [Mass of millicharged DM (MeV)/10] $\times$ 0.115\% $\lesssim f_\mathrm{DM} \lesssim $ 0.4\%.

A large annihilation cross section mediated by new interactions during the freeze-out period is needed to obtain the small fraction of millicharged DM. To explain the 21-cm anomaly and meanwhile to avoid constraints from CMB and 21-cm absorption on s-wave annihilations, the p-wave dominant millicharged DM annihilations during the freeze-out period are considered in Refs. \cite{Berlin:2018sjs,Jia:2018csj,Jia:2018mkc}. Is it possible to explain the 21-cm anomaly with the millicharged DM which being s-wave dominant annihilations during the freeze-out period? Maybe some extraordinary annihilation mechanism could do the job.

For DM s-wave annihilations at the temperature $T \to$ 0, if twice of the DM mass is around the mediator mass, the resonant DM annihilation at $T \to$ 0 would be different from that at the freeze-out period \cite{Ibe:2008ye,Kozaczuk:2015bea,Duch:2017nbe}. Generally, for the mediator mass being slightly below twice of the DM mass, the annihilation cross section of DM at $T \to$ 0 could be larger than that at the freeze-out temperature $T_f$; for the mediator mass being slightly above twice of the DM mass, the annihilation cross section of DM at $T \to$ 0 could be smaller than that at $T = T_f$. In the case of the new mediator mass being slightly above twice of the millicharged DM mass, the millicharged DM with s-wave dominant annihilations may cause the 21-cm anomaly and meanwhile evade constraints from CMB and the 21-cm absorption. This will be investigated in this paper.

\section{Annihilations of millicharged DM}

Which kind of new interactions needed to obtain the small fraction of millicharged DM is an open question. Here we consider the fermionic millicharged DM with dark photon as the new mediator, and now the two mediators are photon and dark photon. The scenario is that: the small fraction of millicharged DM is due to dark photon mediated s-wave annihilations during the freeze-out period, and the 21-cm absorption at the cosmic dawn is caused by photon mediated scattering between millicharged DM and hydrogen. Furthermore, we should keep in mind that there may be more particles in the dark sector, and we focus on the particles that play key roles in transitions/interactions between millicharged DM and ordinary matter.

Besides the fermionic millicharged DM carries a millicharge $\eta e$, here the DM is also dark charged, and dark photon field $\hat{A}'$ mediates dark electromagnetism in the dark sector. The dark photon-photon kinetic mixing $\frac{1}{2} \varepsilon \hat{F}_{\mu\nu} \hat{F}'^{\mu\nu}$ (see e.g., Refs. \cite{Okun:1982xi,Galison:1983pa,Holdom:1985ag,Fayet:1990wx,Foot:2014osa,Bilmis:2015lja,Feng:2015hja,Huang:2018mkk} for more) bridges new transitions between millicharged DM and the standard model (SM) particles, with the field strengths $\hat{F}$ and $\hat{F}'$ corresponding to electromagnetism field $\hat{A}$ and dark electromagnetism field $\hat{A}'$ respectively. The mass of dark photon can be obtained via Higgs-like mechanism or Stueckelberg mechanism \cite{Stueckelberg:1900zz}. After diagonalizing the kinetic mixing with the transformation of $\hat{A} \to A + \varepsilon A'$, $\hat{A}' \to  A'$, the physical eigenstate of dark photon $A'$ couples to SM charged fermions,
\begin{eqnarray}
\mathcal{L}_i^{SM}  = - e \varepsilon  A'_{\mu} J_{\mathrm{em}}^{\mu} ,
\end{eqnarray}
where $J_{\mathrm{em}}^{\mu}$ is the electromagnetic current. In addition, $A'$ couples to the fermionic millicharged DM $\chi$ in forms of $ -  e_D^{} A'_{\mu} \bar{\chi} \gamma^\mu \chi$, where $e_D^{}$ is the dark charge.

For fermionic millicharged DM, the annihilation $\bar{\chi} \chi \to A' \to \mathrm{SM}$ mediated by dark photon $A'$ is an s-wave process, which could be dominant during DM freeze-out. To be able to significantly lower the s-wave annihilation of millicharged DM at low temperature after DM freeze-out, here we consider the case that the mass of dark photon is slightly above twice of the millicharged DM mass. For teens of MeV millicharged DM indicated by the 21-cm absorption, the main annihilation products in SM are $e^+ e^-$, and the annihilation cross section is about
\begin{eqnarray}
\sigma_1 v_r \simeq  \frac{1}{2} \frac{\alpha e_D^2 \varepsilon^2[s(m_\chi^2 + m_e^2) + \frac{s^3}{4m_{A'}^2}](1-\frac{4 m_e^2}{s})^{\frac{1}{2}}}{(s- 2 m_\chi^2)[(s-m_{A'}^2)^2 + m_{A'}^2 \Gamma_{A'}^2]},
\end{eqnarray}
where $v_r$ is the relative velocity of the two DM particles, the factor $\frac{1}{2}$ is for the required $\bar{\chi} \chi$ pair in DM annihilations, and $s$ is the total invariant mass squared. The width $\Gamma_{A'}$ is mainly from $A' \to \bar{\chi} \chi$, with
\begin{eqnarray}
\Gamma_{A'} \approx \frac{e_D^2 (m_{A'}^2 - m_{\chi}^2)}{6 \pi m_{A'}} (1-\frac{4 m_{\chi}^2}{m_{A'}^2})^{\frac{1}{2}} .
\end{eqnarray}
The relic density of millicharged DM $f_\mathrm{DM} \Omega_D h^2$ ($\Omega_D h^2 $ is the total relic density of DM, and $f_\mathrm{DM}$ is the fraction of millicharged DM) is set by the thermally averaged annihilation cross section $\langle \sigma_{1} v_r \rangle$ via the relation \cite{Griest:1990kh,Gondolo:1990dk}
\begin{eqnarray}
f_\mathrm{DM}  \Omega_D h^2  \simeq  \frac{ 1.07 \times 10^9   \mathrm{GeV}^{-1} }{ m_{\mathrm{Pl}}^{}  J_{\mathrm{ann}} \sqrt{g_\ast} }  ,
\end{eqnarray}
with
\begin{eqnarray}
J_{\mathrm{ann}} =  \int^{\infty}_{x_f} \frac{ \langle \sigma_{1} v_r \rangle}{x^2} \mathrm{d} x .
\end{eqnarray}
The parameter $x$ is $x = m_\chi /T$, and $x_f = m_\chi / T_f$ at the freeze-out temperature $T_f$ (see e.g., Ref. \cite{Griest:1990kh} for the calculation of $T_f$). For a pair of DM particles annihilating at $T$ (here $T \ll  m_\chi$), the thermally averaged annihilation cross section can be obtained with methods derived in Ref. \cite{Gondolo:1990dk}. The value of $x_f J_{\mathrm{ann}}$ is a typical annihilation cross section related to the relic abundance of millicharged DM.

For the temperature of DM $T \to$ 0 ($T$ compared with DM mass), the corresponding annihilation cross section of DM mediated by $A'$ is different from that at DM freeze-out period. For $\bar{\chi} \chi \to$ $e^+ e^-$ at $T \to$ 0, contributions from $A'$ and photon are considered, and the annihilation cross section is
\begin{eqnarray}
\sigma_2 v_r &\simeq&  \frac{1}{2} \frac{1}{2 \pi} [(2 m_\chi^2 +m_e^2) (A+B) (A+B)^\ast \\
&& + m_\chi^2 (\frac{4 m_{\chi}^2}{m_{A'}^2}-1)(A A^\ast + 2 B ~ \mathrm{Re} A )] (1 - \frac{m_e^2}{m_{\chi}^2})^{\frac{1}{2}} , \nonumber
\end{eqnarray}
where $A$, $B$ are
\begin{eqnarray}
A =  \frac{e_D \varepsilon e}{4 m_{\chi}^2 - m_{A'}^2 + i m_{A'} \Gamma_{A'}  },  \quad  B =  \frac{\alpha \eta \pi}{m_{\chi}^2 } .   \nonumber
\end{eqnarray}
In addition, the s-wave annihilation mode $\bar{\chi} \chi \to$ $\gamma \gamma$ is deeply suppressed by $\eta^4$.

\section{Numerical analysis}

The millicharged DM is colder than hydrogen at the cosmic dawn. To cool the hydrogen and produce the anomalous 21-cm absorption via photon mediated scatterings between millicharged DM and hydrogens, the parameter ranges for millicharged DM are: the mass $m_{\chi} \sim$ 10$-$35 MeV, the millicharge $\eta e$ with $\eta$ $\sim$ $5 \times 10^{-6} - 5 \times 10^{-5}$, the relic fraction [$m_{\chi}$ (MeV)/10] $\times$ 0.115\% $\lesssim f_\mathrm{DM} \lesssim $ 0.4\%, as given by the Introduction. In the early universe, for $m_{\chi} \sim 10$ MeV, the energy injection from $\bar{\chi} \chi$ annihilations would heat the electron-photon plasma after the electron neutrino decoupling, and this could lower the effective number of relativistic neutrinos $N_{\mathrm{eff}} $. For Dirac fermionic DM, the relation between $N_{\mathrm{eff}}$ and $m_{\chi}/T_d$ was analyzed in Ref. \cite{Ho:2012ug}, with $T_d$ being the neutrino decoupling temperature. Considering the Planck 2018 results \cite{Aghanim:2018eyx} $N_{\mathrm{eff}} = 2.99 \pm 0.17$, we have $m_{\chi}/T_d \gtrsim 5.56$ with $N_{\mathrm{eff}} \gtrsim 2.82$ adopted. Taking $T_d \gtrsim$ 2 MeV, we have $m_{\chi} \gtrsim$ 11.1 MeV. Thus, the mass range of fermionic millicharged DM is 11.1 $ \lesssim m_{\chi} \lesssim$ 35 MeV.

\begin{figure}[htbp!]
\includegraphics[width=0.42\textwidth]{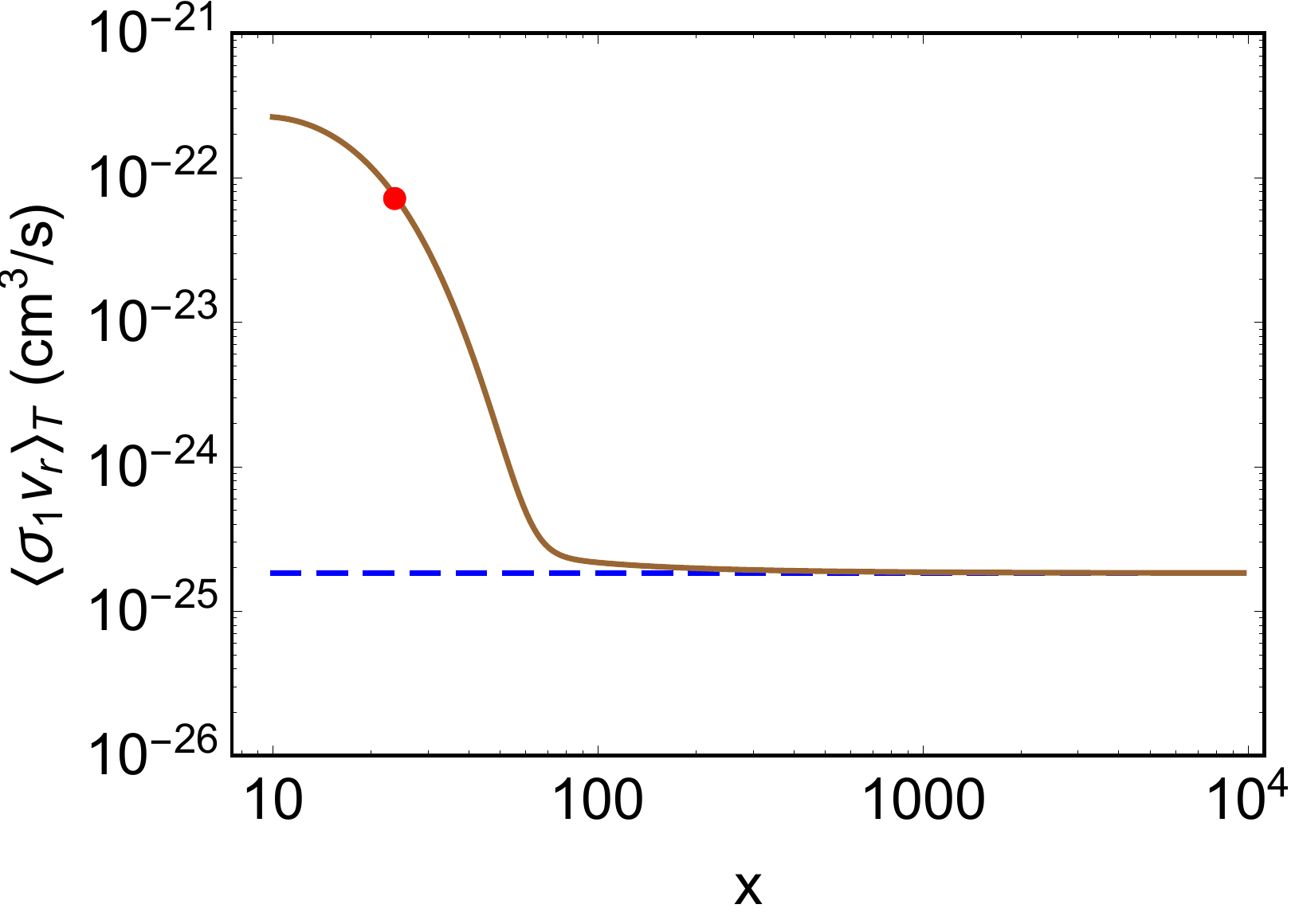} \vspace*{-1ex}
\caption{The temperature-dependent annihilation cross section $\langle \sigma_1 v_r \rangle_{T}$ as a function of $x$, with $x = m_\chi /T$ and $m_\chi =$ 20 MeV. The solid curve is for the case of $f_\mathrm{DM} =$ 0.4\%, $e_D =$ 0.1 and $\xi$ ($\xi = m_{ A'} / 2 m_\chi$) = 1.1. The dot is the annihilation cross section $\langle \sigma_1 v_r \rangle_{T_f}$ at $T = T_f$ ($x_f \simeq$ 23.65). For comparison, the dashed line is the result of $\langle \sigma_1 v_r \rangle_{0}$ at $T \to$ 0.} \label{ann-t}
\end{figure}

\begin{figure}[htbp!]
\includegraphics[width=0.42\textwidth]{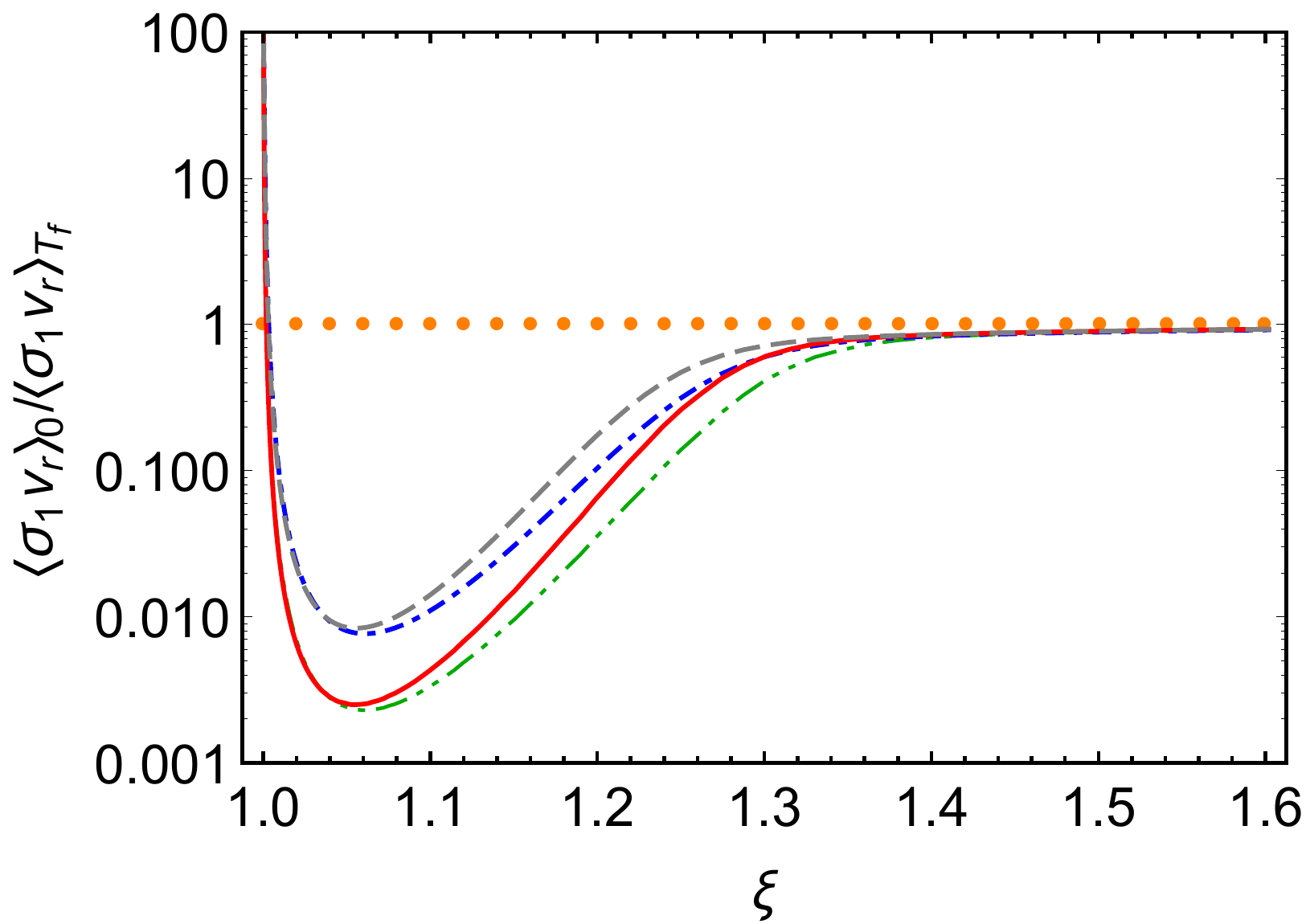} \vspace*{-1ex}
\caption{The ratio $\langle \sigma_1 v_r \rangle_{0}  / \langle \sigma_1 v_r \rangle_{T_f}$ as a function of $\xi$ ($\xi = m_{ A'} / 2 m_\chi$), with $m_\chi =$ 20 MeV. Here the dashed curve is for the case of $\Gamma_{A'} / m_{ A'} = 1 \times 10^{-3}$ and $x_f (x_f = m_\chi/ T_f) =$ 24, and the dot-dashed curve is for the case of $\Gamma_{A'} / m_{ A'} = 1 \times 10^{-3}$ and $x_f =$ 22; the solid curve is for the case of $\Gamma_{A'} / m_{ A'} = 3 \times 10^{-4}$ and $x_f =$ 24, and the dot-dot-dashed curve is for the case of $\Gamma_{A'} / m_{ A'} = 3 \times 10^{-4}$ and $x_f =$ 22. For comparison, the dotted line is for the ratio being equal to 1.}\label{xi-ratio}
\end{figure}

In the Dark Ages, the energy injection from s-wave annihilations of millicharged DM could induce excess heating to the hydrogen gas, and thus the anomalous 21-cm absorption sets stringent constraints on s-wave annihilations of millicharged DM. For the annihilation $\bar{\chi} \chi \to  e^+ e^-$ at $T \to$ 0, if the matter temperature $T_m <$ 4 K is chosen at redshift $z =$ 17.2, the corresponding annihilation cross section is $\lesssim 10^{-26} - 10^{-25} \mathrm{cm}^3/\mathrm{s}$, with $m_{\chi} \sim$ 10$-$35 MeV and $f_\mathrm{DM} =$ 0.01 \cite{Liu:2018uzy}. Thus, to cool the hydrogen and avoid excess heating, the weighted annihilation cross section of $f_\mathrm{DM}^2 \times$ [annihilation cross section] at $T \to$ 0 should be $\lesssim 10^{-30} \mathrm{cm}^3/\mathrm{s}$. Here the s-wave annihilation $\bar{\chi} \chi \to A' \to  e^+ e^-$ is dominant during millicharged DM freeze-out. To escape constraints from CMB and the 21-cm absorption on this s-wave annihilation, we consider the case that the mass $m_{ A'}$ is sightly above $2 m_{\chi}$. Note $\xi = m_{ A'} / 2 m_\chi$, and here $\xi$ is slightly above 1. In this case, the thermally averaged annihilation cross section at temperature $T \to$ 0 could be smaller than that at $T = T_f$. Take $m_\chi =$ 20 MeV, $f_\mathrm{DM} =$ 0.4\%, $e_D =$ 0.1 and $\xi =$ 1.1 as an example to evaluate the temperature-dependent annihilation cross section $\langle \sigma_1 v_r \rangle_{T}$ with $x$ ($x = m_\chi /T$), and the result is shown in Fig. \ref{ann-t}. It can be seen that the corresponding annihilation cross section $\langle \sigma_1 v_r \rangle_{0}$ of millicharged DM at $T \to$ 0 is smaller than $\langle \sigma_1 v_r \rangle_{T_f}$ at the freeze-out period $T = T_f$. To further manifest the resonance effect for different $\xi$, we take $m_\chi =$ 20 MeV, $m_\chi/ T_f = 22, 24$ and $\Gamma_{A'} / m_{ A'} = 1 \times 10^{-3}$, $3 \times 10^{-4}$ as an example to evaluate the ratio of $\langle \sigma_1 v_r \rangle_{0}  / \langle \sigma_1 v_r \rangle_{T_f}$ with $\xi$, and the result is depicted in Fig. \ref{xi-ratio}. It can be seen that, the s-wave annihilation $\bar{\chi} \chi \to A' \to  e^+ e^-$ at $T \to 0$ could be much smaller than that at $T = T_f$, e.g., for $0.02 \lesssim \xi - 1 \lesssim 0.13$, the ratio is $\lesssim 10^{-2}$. Thus, for millicharged DM in MeV scale, the s-wave dominant DM annihilation during the freeze-out period may be allowed by constraints from CMB and the 21-cm absorption, and this will be further analyzed in the following.

\begin{figure}[htbp!]
\includegraphics[width=0.42\textwidth]{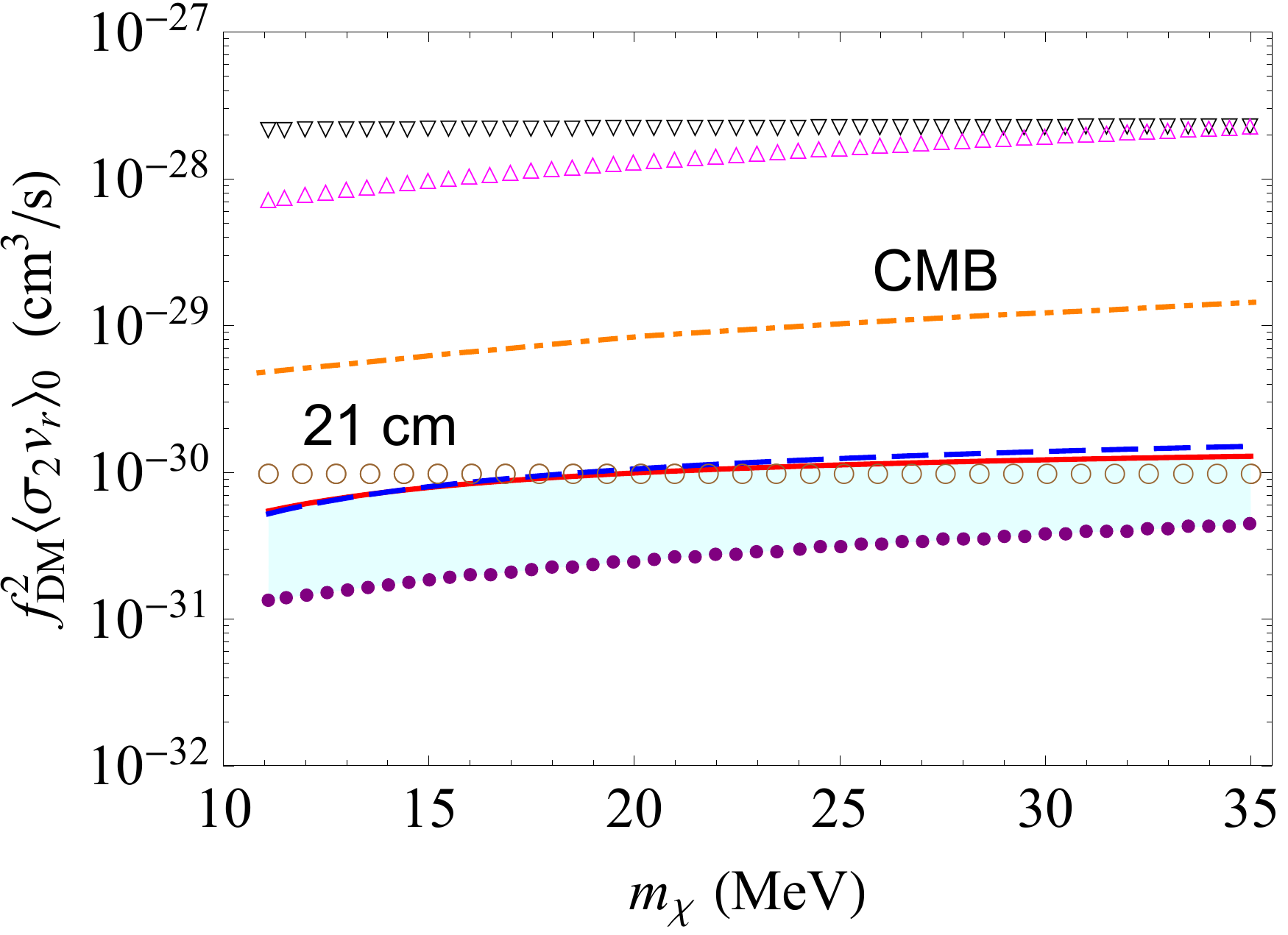} \vspace*{-1ex}
\caption{The weighted annihilation cross section $f_\mathrm{DM}^2 \langle \sigma_{2} v_r \rangle_0$ for given values of $\xi$ ($\xi = m_{ A'} / 2 m_\chi$), with $e_D^{} =$ 0.1 and $m_\chi$ in a range of 11.1$-$35 MeV. The band is the range of $f_\mathrm{DM}^2 \langle \sigma_{2} v_r \rangle_0$, with the parameters $0.004 \lesssim \xi - 1 \lesssim 0.085$, [$m_{\chi}$ (MeV)/10] $\times 0.115\% \lesssim f_\mathrm{DM} \lesssim 0.4\%$, and $5 \times 10^{-6} \lesssim \eta \lesssim  5 \times 10^{-5}$. The upper limit of the band is corresponding to $f_\mathrm{DM} = 0.4\%$, $\eta = 5 \times 10^{-5}$, and $\xi = $ 1.004 (the solid curve), 1.085 (the dashed curve). The solid dotted curve is the lower bound, with $f_\mathrm{DM} =$ [$m_{\chi}$ (MeV)/10] $\times 0.115\%$, $\eta = 5 \times 10^{-6}$, and $\xi \simeq $ 1.0242. The dotdashed curve and empty dotted curve are constraints from the CMB observation \cite{Slatyer:2015jla} and the anomalous 21-cm absorption with $T_m <$ 4 K at $z =$ 17.2 \cite{Liu:2018uzy}, respectively. For comparison, the triangle and reverse triangle curves are the weighted typical annihilation cross section $f_\mathrm{DM}^2 x_f J_{\mathrm{ann}}$ (which is not sensitive to the resonance effect) required to obtain the relic fraction $f_\mathrm{DM} =$ [$m_{\chi}$ (MeV)/10] $\times 0.115\% $ and $f_\mathrm{DM} =$ 0.4\%, respectively.}\label{dm-xi}
\end{figure}

The annihilation cross section of millicharged DM at the freeze-out period is set by the relic density of millicharged DM $f_\mathrm{DM} \Omega_D h^2$, with $\Omega_D h^2 =$ 0.120 $\pm$ 0.001 \cite{Aghanim:2018eyx}. For $\bar{\chi} \chi \to  e^+ e^-$ at $T \to$ 0, suppose the upper limit of the weighted annihilation cross section $f_\mathrm{DM}^2 \langle \sigma_{2} v_r \rangle_0$ (corresponding to the case of $f_\mathrm{DM} \sim $ 0.004 and $\eta \sim  5 \times 10^{-5}$) is tolerated by constraints from CMB and the anomalous 21-cm absorption, and the range of $\xi$ ($\xi = m_{ A'} / 2 m_\chi$) allowed can be derived for a given value of $e_D$ (here $e_D =$ 0.1 is taken), as depicted in Fig. \ref{dm-xi}. It can be seen that, though
constraints of the weighted annihilation cross section $\lesssim 10^{-30}$ $\mathrm{cm}^3/\mathrm{s}$ from the anomalous 21-cm absorption is very strict to additional energy injection from s-wave annihilations of millicharged DM, the s-wave dominant millicharged DM annihilations with 0.004 $\lesssim \xi - 1 \lesssim$ 0.085 can be compatible with the anomalous 21-cm absorption. Thus, the millicharged DM with s-wave dominant annihilations could cool the hydrogen and induce the anomalous 21-cm absorption at the cosmic dawn, and meanwhile avoid excessive energy injection from s-wave annihilations which would cause excess heating to the hydrogen.

\begin{figure}[htbp!]
\includegraphics[width=0.42\textwidth]{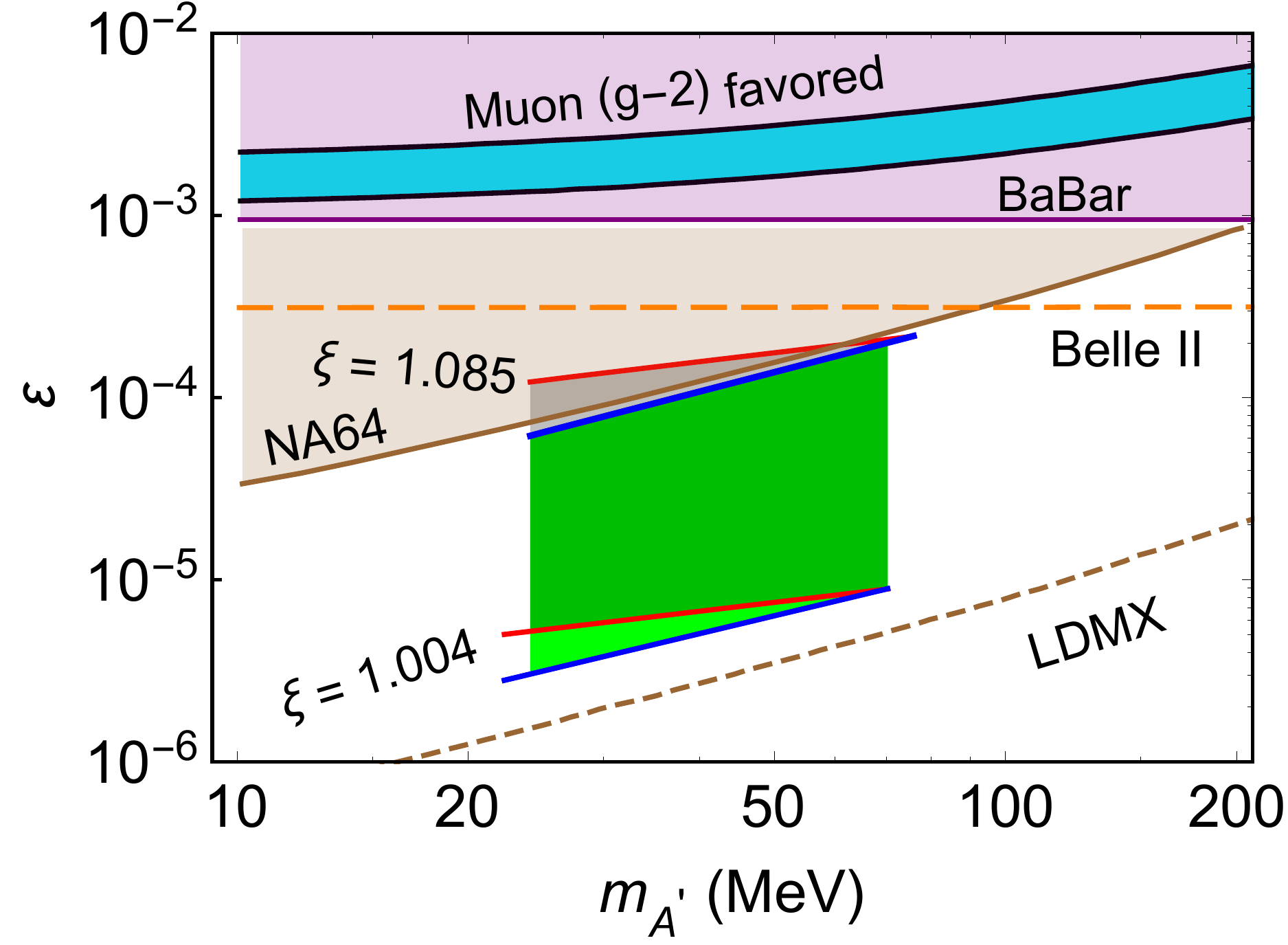} \vspace*{-1ex}
\caption{The value of $\varepsilon$ as a function of $m_{ A'}$ with $\xi$ ($\xi = m_{ A'} / 2 m_\chi$) = 1.085, 1.004, and $e_D^{} =$ 0.1. The bands are ranges of $\varepsilon$ indicated by the 21-cm anomaly. For $\xi =$ 1.085, the upper red, blue solid curves are corresponding to $f_\mathrm{DM} =$ [$m_{\chi}$ (MeV)/10] $\times 0.115\%$, $f_\mathrm{DM} =$ 0.4\%, respectively. For $\xi =$ 1.004, the lower red, blue solid curves are corresponding to $f_\mathrm{DM} =$ [$m_{\chi}$ (MeV)/10] $\times 0.115\%$, $f_\mathrm{DM} =$ 0.4\%, respectively. The constraints from BaBar \cite{Lees:2017lec} and NA64 \cite{NA64:2019imj}, and regions favored by the muon g$-$2 \cite{Bennett:2006fi} are annotated in the figure. The upper and lower dashed curves are the expected sensitivity set by 20 fb$^{-1}$ Belle II data \cite{Kou:2018nap} and the ultimate reach of LDMX \cite{Akesson:2018vlm}, respectively.}\label{darkp-m}
\end{figure}

The dark photon mainly decays into $\bar{\chi} \chi$, and this invisible decay could be produced at lepton collider and beam dump experiments \cite{Izaguirre:2013uxa,Banerjee:2016tad,Banerjee:2017hhz,NA64:2019imj,Lees:2017lec}, or the kinetic mixing parameter $\varepsilon$ would be restricted by experiments. For a given $\xi$ ($\xi = m_{ A'} / 2 m_\chi$), to obtain the small fraction $f_\mathrm{DM}$ of millicharged DM, the range of $\varepsilon$ is derived with $\xi =$ 1.085, 1.004, and $e_D^{} =$ 0.1, as shown in Fig. \ref{darkp-m}. It can be seen that the range of $\varepsilon$ indicated by the 21-cm absorption is allowed by recent lepton collision experiments, such as BaBar \cite{Lees:2017lec} and NA64 \cite{Banerjee:2017hhz,NA64:2019imj}. The dark photon can be further investigated at future experiments, such as NA64 \cite{NA64:2019imj}, Belle II \cite{Kou:2018nap} and Light Dark Matter eXperiment (LDMX) \cite{Akesson:2018vlm}.

Now we give a brief discussion about the detection of millicharged DM at underground experiments. For millicharged DM of concern, magnetic fields in the Milky Way could expel most of millicharged DM from the Galactic disk, as estimated in Refs. \cite{Barkana:2018lgd,Chuzhoy:2008zy,McDermott:2010pa}. Even though a small amount of millicharged DM are remained in the Galactic disk, the magnetic fields related to the solar wind and the Earth's magnetic field would substantially reduce the flux of millicharged DM arriving to the Earth's surface. In addition, for underground experiments, the terrestrial effect of a particle penetrating the earth and strongly interacting with overburden matter (e.g., photon/dark photon mediated large interactions related to the electric charge of nucleus) could deplete the particle's energy and significantly reduce the detection sensitivity \cite{Emken:2017erx,Emken:2019tni}. For the millicharged DM, the reference cross section $\bar{\sigma}_e$ (see, e.g. Ref. \cite{Essig:2011nj} for more) of $\chi -$electron scattering with photon as the mediator is in a range of $\sim$ 3.5 $\times 10^{-26} -$ 3.5 $\times 10^{-24}$ cm$^2$, and the rock/concrete shielding with depths of $\sim 3-10$ meters could result in little detection signal of millicharged DM \cite{Emken:2019tni}. In this case, the millicharged DM of concern will evade constraints from underground experiments, such as XENON10 \cite{Essig:2012yx,Essig:2017kqs}, XENON100 \cite{Essig:2017kqs}, and DarkSide-50 \cite{Agnes:2018oej}. Moreover, the above case may be not the whole thing for the millicharged DM, as analyzed in Ref. \cite{Dunsky:2018mqs}. The millicharged DM could be accelerated by supernova shocks, and the evacuation of millicharged DM from the disk may not be effective due to the diffusion of millicharged DM from the halo \cite{Dunsky:2018mqs}. Hence, there are uncertainties about the millicharged DM in the disk, and corresponding uncertainties in direct detections.

\section{Conclusion and discussion}

In this paper, the s-wave dominant annihilations of MeV millicharged DM has been studied with the anomalous 21-cm absorption. The photon mediated scattering could cool the hydrogen and induce the 21-cm anomaly at the cosmic dawn, and the required small fraction $f_\mathrm{DM}$ of millicharged DM is predominantly contributed by the dark photon mediated annihilations during the freeze-out period. For s-wave dominant DM annihilations, to be compatible with stringent constraints from CMB and the anomalous 21-cm absorption, the annihilation is considered being close to the resonance and $\xi$ ($\xi = m_{ A'} / 2 m_\chi$) being slightly above 1. In this case, the annihilation cross section at $T \to 0$ could be much smaller than that at $T = T_f$. For $\xi$ in a range of $0.004 \lesssim \xi - 1 \lesssim 0.085$ ($e_D^{} =$ 0.1), the weighted annihilation cross section $f_\mathrm{DM}^2 \langle \sigma_{2} v_r \rangle_0$ could be $\lesssim 10^{-30}$ $\mathrm{cm}^3/\mathrm{s}$, which is tolerated by constraints from the anomalous 21-cm absorption with $T_m <$ 4 K ($z =$ 17.2), avoiding excess heating to the hydrogen.

For millicharged DM with the millicharge $\eta e$, the spatial magnetic fields and the terrestrial effect of large interactions between DM and ordinary matter result in the low-velocity millicharged DM remained in the disk evading DM direct detection experiments, while the millicharged DM accelerated by supernova shocks may be detectable. As there are uncertainties about the millicharged DM in the disk, the corresponding further explorations of millicharged DM are needed. The beam dump and lepton collider experiments can do the job of hunting for millicharged DM, such as NA64, Belle II and LDMX, especially for the case of a large $\xi -1$. We look forward to the exploration of millicharged DM at future lepton experiments. In addition, neutrino experiments could also be employed to search for MeV DM \cite{Ge:2017mcq,DeRomeri:2019kic}, and the terrestrial effect is needed to be taken into account for the investigation of millicharged DM.

\acknowledgments \vspace*{-3ex} This work was partly supported by National Natural Science Foundation of China under the contract No. 11505144, and Longshan Academic Talent Research Supporting Program of SWUST under the contract No. 18LZX415.

\end{document}